\documentclass[prb,twocolumn,amsmath,amssymb]{revtex4-1}

\usepackage{graphicx}
\usepackage{dcolumn}
\usepackage{bm}

\begin{document}

\title{Multi-component magneto-optical conductivity of multilayer graphene on SiC}

\author{I. Crassee$^{1}$, J. Levallois$^{1}$, D. van der Marel$^{1}$, A. L. Walter$^{2,3}$,
Th. Seyller$^{4}$ and A. B. Kuzmenko$^{1}$}

\affiliation{
\\
\mbox{ }
\\
$^{1}$D\'epartement de Physique de la Mati\`ere Condens\'ee,
Universit\'e de Gen\`eve, CH-1211 Gen\`eve 4,
Switzerland\\
\\
$^{2}$Dept. of Molecular Physics, Fritz-Haber-Institut der
Max-Planck-Gesellschaft, Faradayweg 4-6, 14195 Berlin,
Germany\\
\\
 $^{3}$E. O. Lawrence Berkeley National Laboratory, Advanced
Light Source, MS6-2100, Berkeley, CA 94720\\
\\
$^{4}$Lehrstuhl f\"{u}r Technische Physik, Universit\"{a}t
Erlangen-N\"{u}rnberg, Erwin-Rommel-Str. 1, 91058 Erlangen,
Germany }


\date{\today}

\begin{abstract}
Far-infrared diagonal and Hall conductivities of multilayer
epitaxial graphene on the C-face of SiC were measured using
magneto-optical absorption and Faraday rotation in magnetic
fields up to 7 T and temperatures between 5 and 300 K. Multiple
components are identified in the spectra, which include: (i) a
quasi-classical cyclotron resonance (CR), originating from the
highly doped graphene layer closest to SiC, (ii) transitions
between low-index Landau levels (LLs), which stem from weakly
doped layers and (iii) a broad optical absorption background.
Electron and hole type LL transitions are optically
distinguished and shown to coexist. An electron-hole asymmetry
of the Fermi velocity of about 2\% was found within one
graphene layer, while the Fermi velocity varies by about 10\%
across the layers. The optical intensity of the LL transitions
is several times smaller than what is theoretically expected
for isolated graphene monolayers without electron-electron and
electron-phonon interactions.

\end{abstract}

\maketitle

\section{Introduction}

Graphene attracts much interest due to the intriguing physical
properties and a potential for novel applications. Epitaxially
grown graphene on SiC \cite{BergerJPCB04}  is particularly
promising for large-scale production. The growth conditions of
epitaxial graphene were constantly improving in the last years,
which resulted in a macroscopic continuity of graphene layers
and an enhanced mobility of charge carriers
\cite{BergerJPCB04,BergerScience06,EmtsevNatureMat09,RiedlPRL09,
SpeckMatSciForum10,TzalenchukNatureNano10}. Electronic
properties of epitaxial graphene grown on Si- and C-faces of
SiC, are markedly different \cite{HassPRL08,FirstMRSBull10}.
Interestingly, multilayer epitaxial graphene produced on the
carbon face shows a number of electronic features typical of
isolated monolayer graphene as revealed by infrared
spectroscopy \cite{SadowskiPRL06}, scanning tunneling
microscopy (STM) \cite{MillerScience09,SongNat10},
angle-resolved photoemission spectroscopy (ARPES)
\cite{SprinklePRL09} and quantum Hall effect \cite{WuAPL09}. In
particular, the Landau levels in this material demonstrate a
square-root dependence on the perpendicular magnetic field $B$
and the level index $n ( = 0, \pm 1...)$:

\begin{equation}
E_n=E_D + \mbox{sgn}(n) \sqrt{2 e \hbar v_F^2 |nB|},
\label{EqEn}
\end{equation}

\noindent where $v_F$ is the Fermi velocity and $E_D$ is the
Dirac-point energy. It is generally believed that such an
effective electronic interlayer decoupling is a result of
twisted, non-Bernal stacking of the C-face grown graphene
layers. Significant progress in theoretical understanding of
the influence of stacking on the electronic structure and
Landau levels was made \cite{GuineaPRB06, mcCannPRL06,
LopesDosSantosPRL07, ShallcrossPRL08, LaissardiereNanoLett10,
ShallcrossPRB10, MelePRB01, BistritzerPRB10,
BistritzerArxiv11}. However, no complete theory able to
quantitatively predict the effect of twisting on the band
structure at an arbitrary stacking angle exists at the moment
even for bilayer graphene. In real samples, many layers are
present with a random rotation between each pair of neighbors.
Moreover, the substrate induces a strong variation of the
Dirac-point energy $E_D$ with respect to the chemical potential
and therefore a very different density and mobility of carriers
in different layers. More experiments are needed to understand
the complex electronic structure of this system and establish
favorable conditions for applications.

Infrared spectroscopy, which is a direct probe of charge
dynamics, is well suited to study carriers in graphene. Optical
spectra contain contributions from all graphene layers,
including those that are not seen by surface probes, such as
ARPES and STM. In a recent work \cite{CrasseeNatPhys11}, we
measured the rotation of polarization of light passing through
epitaxial graphene in a magnetic field, known as the Faraday
effect. Being an optical analogue to the d.c. Hall effect, the
Faraday rotation reveals the sign of the charge carriers
involved in various LL transitions. Moreover, optical
spectroscopy allows separating carriers by their high-frequency
dynamics, in contrast to the d.c. transport measurements. Here
we present an extensive magneto-optical study of multilayer
epitaxial graphene grown on the carbon side of SiC, by
combining the Faraday rotation measurements with the
transmission spectra and extracting both the diagonal and Hall
optical conductivity. Different magneto-optical contributions
are disentangled and studied quantitatively using a
multi-component CR model. This approach reveals a number of
interesting properties of multiple charge carriers in
multilayer graphene, for example, the simultaneous presence of
both electron- and hole-like components as well as the
existence of carriers with different Fermi velocities. It also
allows us to analyze optical intensities of various transitions
and make comparison to theoretical models. Finally, the
dependence of optical spectra on temperature and environmental
(surface) doping are studied.

\section{Techniques}\label{magneto-optical experiment}

The rotationally stacked multilayer graphene on the carbon
terminated side of 6H-SiC was produced by hydrogen etching at
1450$^{o}$C and subsequent graphitization in an argon
atmosphere at 1650$^{o}$C. The substrate has a thickness of 370
$\mu$m and a surface area of 10$\times$10 mm$^{2}$. The back
side of the substrate was cleaned from undesirably grown
graphene using scotch tape. Using X-ray photoemission
spectroscopy we estimated the number of graphene layers to be 5
$\pm$ 1 and checked that the backside is graphene free.
Additionally, we performed mid-infrared transmission microscopy
over an area comparable to the total area used for the
magneto-optical experiment using a spot size of 10$\times$10
$\mu$m$^{2}$. We assume that each graphene layer absorbs 1.5\%
of light at 1 eV, which is the value expected for monolayer
graphene on SiC, provided that the chemical potential does not
exceed 0.5 eV. This absorption value is lower than the
absorption of 2.3\% of free-standing graphene, because the
refractive index of SiC is larger than 1. The infrared
absorption shows a distribution peaked at 7 layers with a FWHM
of 2.7 layers. The slightly larger number of layers revealed by
infrared microscopy can be related to an underestimation of the
absorption of one layer. Notably, both measurements show that
at least 4 layers are present at any point of the sample.

We measured the Faraday rotation $\theta(\omega)$ and
magneto-optical transmission $T(\omega)$ in the range of photon
energies $\hbar\omega$ between 8 and 80 meV as described in
Ref. \onlinecite{CrasseeNatPhys11}. The optical transmission
was measured with respect to a bare SiC substrate without
graphene, which underwent similar hydrogen etching at
1450$^{o}$C as the graphitized sample. The optical spot had a
diameter of 5 mm. Two series of measurements were performed:
first by varying temperature between 5 and 300 K at a fixed
magnetic field of 3 T and then by varying magnetic field from 0
to 7 T at a constant temperature of 5 K. Within each series,
the optical absorption and the Faraday rotation were measured
one after another with a shortest possible delay. During the
experiments the sample was in a He gas flow; between the
production and the experiments, as well as between the first
and second series, it was stored in desiccated air. As will be
shown later, even these precautions did not allow us to
completely avoid surface contamination, which has an effect on
charge dynamics in the top layers. Therefore, the data
presented in this paper differ somewhat from data presented in
Ref. \cite{CrasseeNatPhys11}, although the same sample was
used.

The real parts of the diagonal, $\sigma_{xx}(\omega)$, and
Hall, $\sigma_{xy}(\omega)$, optical conductivities can be
directly obtained from the magneto-optical absorption and
Faraday rotation by using the general thin-film approximation,
taking internal reflections in the substrate into account:
\begin{eqnarray}
1-T(\omega) &\approx& 2 Z_0 f_s(\omega)
\mbox{Re}[\sigma_{xx}(\omega)]
\label{EqThetaSxx}\\
\theta(\omega) &\approx& Z_0 f_s(\omega)
\mbox{Re}[\sigma_{xy}(\omega)]\label{EqThetaSxy}
\end{eqnarray}
\noindent where $Z_0$ $\approx$ 377 $\Omega$ is the impedance
of vacuum and $f_{s}(\omega)$ is a spectrally smooth
dimensionless function specific to the
substrate\cite{CrasseeNatPhys11}. These relations are obtained
by linear expansion of the exact Fresnel formulas, which is
accurate for our sample with only a few graphene layers. We
reduced the spectral resolution to 1 meV in order to suppress
the Fabry-Perot interference in the substrate. In this case:
\begin{equation}
f_s(\omega) = \frac{1}{n_s +
1}+\frac{2n_s}{n_{s}^2-1}\cdot\frac{q^2}{1-q^2}
\label{Eqfs}
\end{equation}
\noindent where $q = \left[(n_s - 1)/(n_s + 1)\right]^2
\exp\left[-(2\omega/c)k_s d\right]$, $d$, $n_s(\omega)$ and
$k_s(\omega)$ are the thickness, the refractive index and
extinction coefficient of the substrate. The latter quantities,
which are independent of magnetic field, were determined from
measurements of the absolute transmission and reflection
spectra of the bare SiC at every used temperature.

The complex magneto-optical conductivity tensor:
\begin{equation}\label{condtensor}
\hat{\sigma}(\omega)=\left[
\begin{matrix}
\sigma_{xx}(\omega) & \sigma_{xy}(\omega) \\
-\sigma_{xy}(\omega) & \sigma_{xx}(\omega)
\end{matrix}\right]
\end{equation}
\noindent is diagonal in the circular basis $\vec{x}\pm
i\vec{y}$, with eigenvalues:
\begin{equation}
\sigma_{\pm}(\omega)= \sigma_{xx}(\omega)\pm i
\sigma_{xy}(\omega). \label{EqSpm}
\end{equation}
\noindent The absorption of right- and left-handed circular
polarizations (which are shown in the insets of
Fig.\ref{Figcomponents}c and \ref{Figcomponents}d) is described
by the real parts of $\sigma_{+}(\omega)$ and
$\sigma_{-}(\omega)$ respectively. This basis provides an
intuitive way to present the magneto-optical conductivity
showing individual LL transitions, which are active for
strictly one circular polarization or the other
\cite{AbergelPRB07,GusyninJPCM07,OrlitaSemiSciTech10}. In
particular, for right circularly polarized light, transitions
between electron like LLs with $n\geq0$, such as
LL$_{0\rightarrow1}$, LL$_{1\rightarrow2}$ etc, are active,
provided that the chemical potential is in between the
corresponding LLs (the symbol LL$_{i\rightarrow j}$ is used to
designate the LL transition between levels $i$ and $j$).
Similarly, for left circularly polarized light, transitions
between hole like LLs, (LL$_{-1\rightarrow 0}$,
LL$_{-2\rightarrow-1}$, etc.) are excited \cite{text}.

From Eq. (\ref{EqSpm}) one can see that the imaginary part of
$\sigma_{xy}(\omega)$ is needed to obtain the real part of the
magneto-optical conductivity in the circular basis.
Experimentally, $\mbox{Im}[\sigma_{xy}(\omega)]$ is directly
related to the ellipticity of the transmitted light. However,
in the present experiment the error bars on the ellipticity are
larger than the ones on the Faraday angle. Therefore, instead
of directly applying  Eq. (\ref{EqSpm}) to the experimental
data, we plot $\sigma_{\pm}(\omega)$ obtained from the results
of the multi-component fitting discussed in Section \ref{multi
component fitting}. For simplicity, in the rest of the paper,
the symbols $\sigma_{xx}$, $\sigma_{xy}$, $\sigma_{+}$ and
$\sigma_{-}$ will be used to denote the real parts of the
corresponding complex functions, unless stated otherwise.

\section{Results}

\subsection{Magneto-optical conductivity at 3 T and 5 K}\label{multi component
fitting}
\begin{figure*}
\includegraphics[width=16cm]{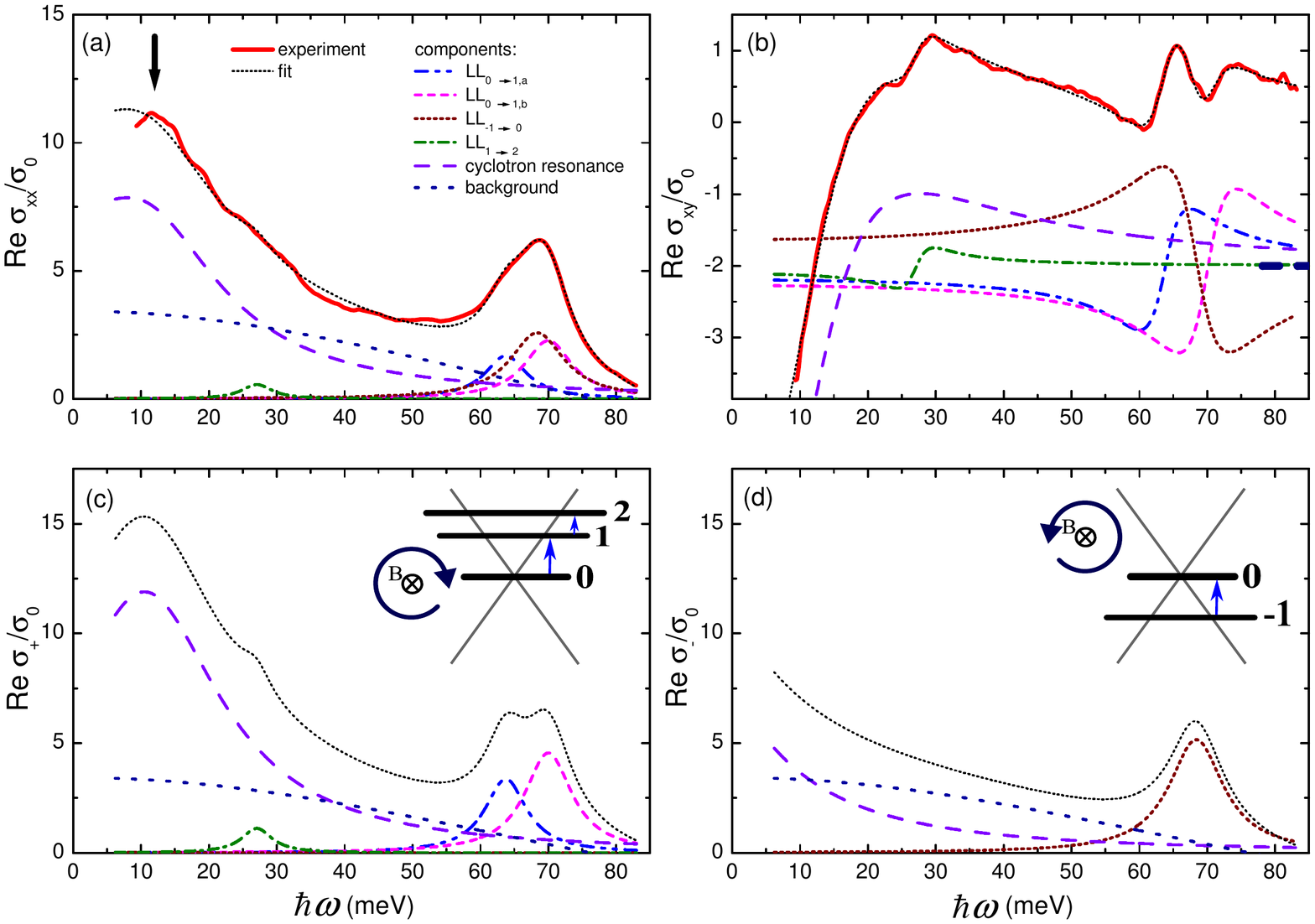}\\
\caption{The diagonal (a) and Hall (b) optical conductivity at
3 T and 5 K, normalized to the universal conductivity
$\sigma_0$, and results of the multi-component modeling
described in the text. The components in the graph are the
individual contributions needed to create the total fit shown
by the black dotted line on top of the data. The components in
panel (b) have an offset indicated by the blue dashed line at
the right side of the graph. Panels (c) and (d): the optical
conductivities in the circular basis $\sigma_{+}(\omega)$ and
$\sigma_{-}(\omega)$ corresponding to the absorption of right
and left circular polarized light respectively. In the insets
the low-index LL transitions allowed in the corresponding
polarizations are shown.}\label{Figcomponents}
\end{figure*}

Figure \ref{Figcomponents} shows representative conductivity
spectra at $B = $ 3 T and $T = $ 5 K. The diagonal and Hall
conductivities are plotted as solid lines in Fig.
\ref{Figcomponents}a and \ref{Figcomponents}b. They are
expressed in dimensionless units of $\sigma_{0} =
e^{2}/4\hbar$, which is equal to the universal optical
conductivity of monolayer graphene \cite{AndoJPSJ02}. The rich
structure of the spectra indicates the presence of multiple
optical transitions. The absorption at low energy, shown by the
arrow, corresponds to quasi-classical CR coming from the highly
doped graphene layer closest to the SiC
substrate\cite{CrasseeNatPhys11}. At about 60 to 70 meV, a
strong peak in $\sigma_{xx}(\omega)$ is observed that matches
the energy of the LL$_{0\rightarrow1}$ or LL$_{-1\rightarrow0}$
transitions at this field. The Hall conductivity displays a
`zig-zag' shape in this spectral range, suggesting that there
are multiple components contributing to the optical response.
In addition, a structure is present at about 27 meV, most
clearly seen in $\sigma_{xy}(\omega)$, resulting from the
LL$_{1\rightarrow2}$ transition.

In order to disentangle different contributions to the
magneto-optical spectra we used a multi-component model where
the total conductivity is given by a sum of separate cyclotron
resonances. In the circular basis the total complex
magneto-optical response is given by:
\begin{eqnarray}
\sigma_{\pm}(\omega) & = &\sum_{j}\frac{2W_j}{\pi} \cdot
\frac{i}{\omega \mp \omega_{c,j}+i\gamma_j} \label{EqSpm2}
\end{eqnarray}
\noindent and the complex diagonal and Hall optical
conductivities are accordingly expressed by:
\begin{eqnarray}
\sigma_{xx}(\omega) & = &\sum_{j}\frac{2W_j}{\pi} \cdot
\frac{\gamma_j-i\omega}{\omega_{c,j}^2-(\omega+i\gamma_j)^2} \label{EqSxx}\\
\sigma_{xy}(\omega) & = & \sum_{j}\frac{2W_j}{\pi} \cdot
\frac{-\omega_{c,j}}{\omega_{c,j}^2-(\omega+i\gamma_j)^2},
\label{EqSxy}
\end{eqnarray}

\noindent where $\omega_{c,j}$ is the cyclotron frequency,
$W_j$ is the spectral weight  and $\gamma_{j}$ is the
broadening of the $j$th component. Physically speaking, each
component can describe either a quasi-classical CR or a
transition between individual LLs separated by energy
$|\hbar\omega_{c,j}|$. Notably, $\sigma_{xx}(\omega)$ does not
depend on the sign of the charge carriers, unlike
$\sigma_{xy}(\omega)$, where the sign can be derived directly
from the spectral shape\cite{CrasseeNatPhys11}. In
$\sigma_{xx}(\omega)$ each component results in a peak centered
at $|\omega_{c,j}|$. In $\sigma_{xy}(\omega)$ however, it shows
an antisymmetric structure, where $|\omega_{c,j}|$ corresponds
to the inflection point in the curve. The slope at this point
coincides with the sign of the cyclotron frequency and reveals
the polarity of the charge carriers involved in the
transition\cite{CrasseeNatPhys11}. In particular, a positive
(negative) slope signals electron (hole) like carriers.

The spectra $\sigma_{xx}(\omega)$ and $\sigma_{xy}(\omega)$
were simultaneously fitted using Eqs. (\ref{EqSxx}) and
(\ref{EqSxy}), while allowing the parameters $\omega_{c,j}$,
$W_j$ and $\gamma_{j}$ to change freely. The fitting curves are
shown in Fig. \ref{Figcomponents}a and \ref{Figcomponents}b, as
black dotted lines. We found that a minimal model describing
satisfactorily the spectral structures contains six components,
which are shown separately in the same panels.

Interestingly, it was necessary to introduce at least three
components with transition energies between 60 meV and 70 meV
to describe the structure in this range. Two of the resonances
are electron like, which we designate as LL$_{0\rightarrow1,a}$
and LL$_{0\rightarrow1,b}$, and one is hole like, referred to
as LL$_{-1\rightarrow0}$. Note that all these transitions are
actually at different energies, which is clearly seen already
from the presence of three inflection points in the optical
Hall conductivity. A fourth component is at about 27 meV and
corresponds to the electron like LL$_{1\rightarrow2}$
transition. The fifth component with a small value of
$\hbar\omega_c$ = 9 meV originates from a quasi-classical
electron like CR. Finally, there is a component with zero
cyclotron frequency and large scattering, which forms a broad
absorption background present in $\sigma_{xx}(\omega)$, but
absent in $\sigma_{xy}(\omega)$. A possible origin of this
background will be discussed in Section \ref{Discussion}.

The different contributions from electrons and holes to the
magneto-optical conductivity are easily seen in the basis of
circularly polarized light as shown in Fig.
\ref{Figcomponents}c and \ref{Figcomponents}d. The electron
like transitions LL$_{0\rightarrow1,a}$,
LL$_{0\rightarrow1,b}$, LL$_{1\rightarrow2}$ as well as the CR
show peaks in $\sigma_{+}(\omega)$, while the hole like
LL$_{-1\rightarrow0}$ transition manifests itself in
$\sigma_{-}(\omega)$. Because of the small value of $\omega_c$
and a relatively large scattering, the CR component has also a
tail in $\sigma_{-}(\omega)$. The absorption background
contributes equally to $\sigma_{+}(\omega)$ and
$\sigma_{-}(\omega)$.

\subsection{Dependence on magnetic field}\label{magnetic field dependence}

\begin{figure*}\tt
\includegraphics[width=17.5cm]{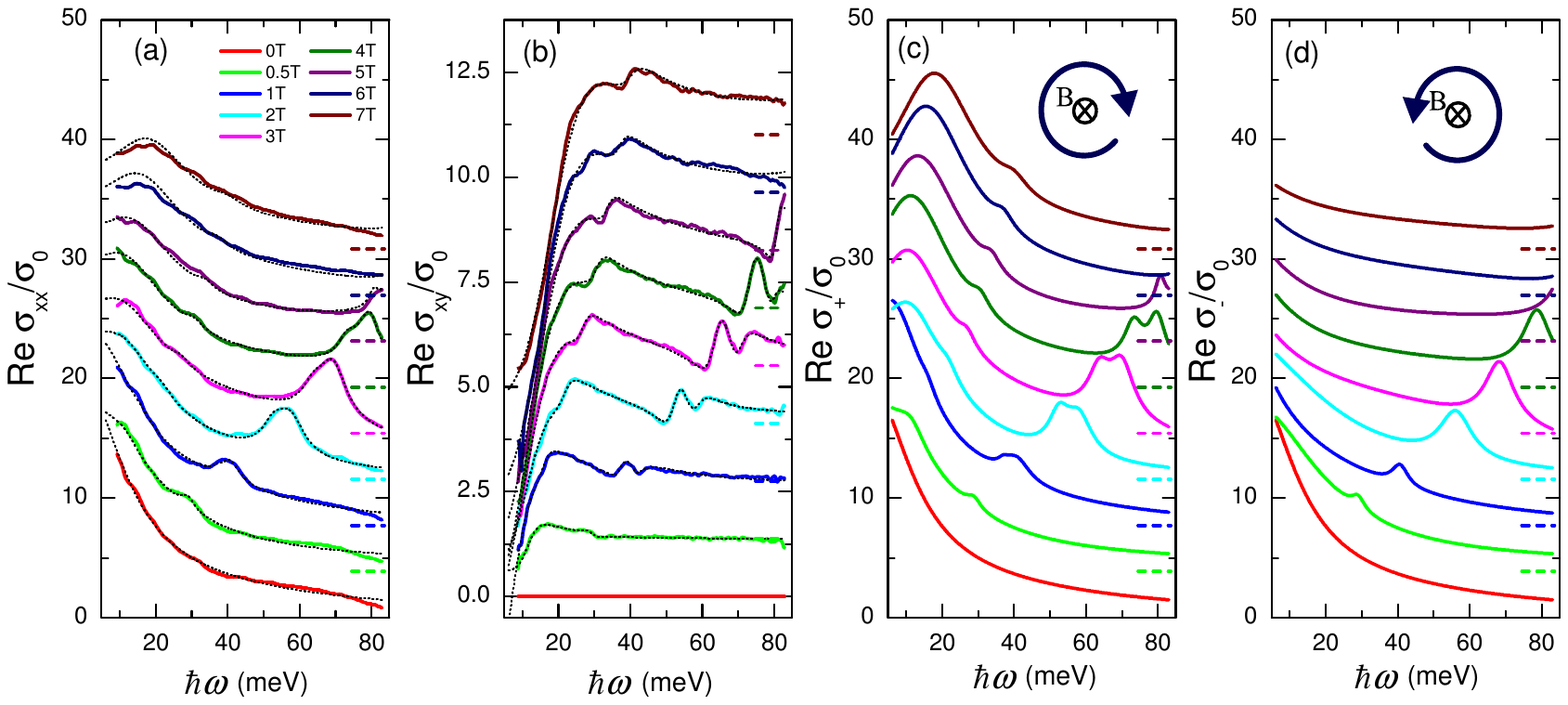}\\
\caption{Magneto-optical conductivity of multilayer graphene at
5 K, normalized to the universal conductivity $\sigma_0$, for
several magnetic fields up to 7 T. The curves in all panels are
offset as indicated by the dashed lines. Larger offsets
correspond to higher magnetic fields. Panels (a) and (b) show
the measured spectra of $\sigma_{xx}(\omega)$ and
$\sigma_{xy}(\omega)$ (solid lines) and multi-component fits
(black dotted lines). In panels (c) and (d) the model-derived
$\sigma_{+}(\omega)$ and $\sigma_{-}(\omega)$ are shown.}
\label{Figmagneticfielddependence}
\end{figure*}

\begin{figure}
\includegraphics[width=8.5cm]{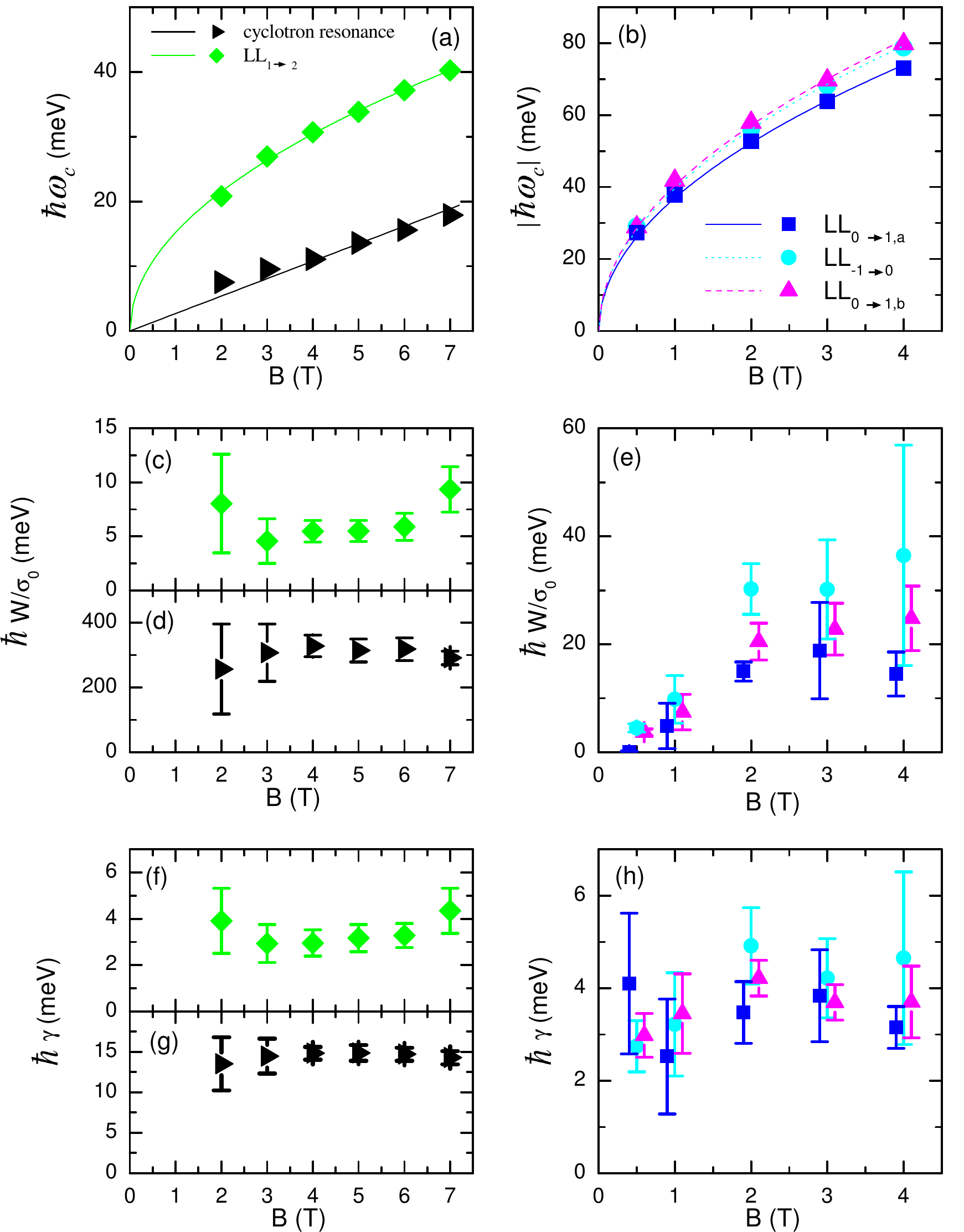}\\
\caption{Parameters of various components as a function of magnetic field
obtained from the multi-component fits to $\sigma_{xx}(\omega)$
and $\sigma_{xy}(\omega)$ of Fig.\ref{Figmagneticfielddependence}.
(a) The transition energies of the CR (triangles) and the
LL$_{1\rightarrow2}$ transition (diamonds). The solid lines are fits using Eq. (\ref{Eqwcef}) and Eq. (\ref{EqEn}) respectively.
(b) The transition energies of LL$_{0\rightarrow1,a}$ (squares),
LL$_{0\rightarrow1,b}$ (triangles) and LL$_{-1\rightarrow0}$ (circles) transitions.
The solid lines are fits using Eq. (\ref{EqEn}). (c) Spectral weight
of the LL$_{1\rightarrow2}$ transition. (d) Spectral weight
of the CR peak. (e) Spectral weights of the
LL$_{0 \rightarrow1,a}$, LL$_{0\rightarrow1,b}$ and the LL$_{-1
\rightarrow0}$ transitions. (f) The broadening of the LL$_{1
\rightarrow2}$ transition. (g) The broadening of the CR
transition. (h) The broadening of LL$_{0
\rightarrow1,a}$, LL$_{0\rightarrow1,b}$ and the $-1
\rightarrow 0$ transitions. Errorbars in all figures are the
standard deviation of several fitting results. The error bars in panels (a) and (b)
are within the symbol size.
} \label{Figfitresults2nd}
\end{figure}

The magnetic field dependence of $\sigma_{xx}(\omega)$ and
$\sigma_{xy}(\omega)$, measured at 5 K, is shown in Fig.
\ref{Figmagneticfielddependence}a and
\ref{Figmagneticfielddependence}b. The multi-component
character of the spectra is present at different fields,
although the resonances show a strong field dependence. The
model described in Section \ref{multi component fitting} was
used to fit $\sigma_{xx}(\omega)$ and $\sigma_{xy}(\omega)$
simultaneously for every field. The fits, shown as black dotted
lines, are used to calculate the magneto-optical conductivities
in the basis of right and left handed circularly polarized
light according to Eq. (\ref{EqSpm2}), which are plotted in
Fig. \ref{Figmagneticfielddependence}c and
\ref{Figmagneticfielddependence}d.

The field dependence of all fitting parameters is shown in Fig.
\ref{Figfitresults2nd}. The energy of the CR is linear in
field, as seen in Fig. \ref{Figfitresults2nd}a. This
quasi-classical behavior is expected for highly doped graphene,
where large-index LLs are close to the Fermi energy
\cite{AndoJPSJ02, GusyninNJP09}. In this case the
quasi-classical cyclotron frequency is inversely proportional
to the Fermi energy $\epsilon_F$:

\begin{equation}
\omega_c=\frac{|e|Bv_F^2}{\epsilon_F}. \label{Eqwcef}
\end{equation}

\noindent Using this relation and taking as an estimate $v_F =
1.0 \times 10^{6}$ m/s we obtain $\epsilon_F$ = 0.24 eV, which
corresponds to a carrier concentration $ n =
 \epsilon_F^2 / \pi v_F^2 \hbar^2 =$ 4.2 $\times$
10$^{12}$ cm$^{-2}$.

The spectral weight $W$ of the cyclotron peak, shown in Fig.
\ref{Figfitresults2nd}d, is field independent within the
experimental accuracy. In the absence of interactions, the
Drude weight in a single graphene layer is related to
$\epsilon_F$ by the formula:

\begin{equation}
\frac{\hbar W}{\sigma_0} = 2|\epsilon_F|. \label{EqWef}
\end{equation}

\noindent This gives $\epsilon_F$ = 0.16 eV  and accordingly $n
=$ 1.9 $\times$ 10$^{12}$ cm$^{-2}$, which are significantly
smaller than the values based on the cyclotron frequency. It is
possible that interactions renormalize the Drude weight and
spread the missing weight over a large spectral range. In our
fits this missing weight might be `absorbed' by the broad
background component mentioned above. On the other hand, we
obtain a reduced Drude weight based on the assumption that for
the bottom layer $v_F = 1.0 \times 10^{6}$ m/s. Accepting a
smaller Fermi velocity (by about 20\%) would make the cyclotron
resonance and the Drude weight match according to Eqs.
(\ref{Eqwcef}) and (\ref{EqWef}).

The scattering rate $\hbar \gamma$ (Fig.
\ref{Figfitresults2nd}g) of the cyclotron peak is field
independent and equal to about 14 meV. Using the semi-classical
relation:

\begin{equation}
\mu = \frac{|\omega_c|}{\gamma |B|} \label{Eqmu}
\end{equation}

\noindent we find a mobility of about 2000 cm$^{2}$/Vs, based
on the experimental values of the scattering rate and the
cyclotron frequency. Magneto-transport measurements show that
the carrier density and mobility in multilayer graphene are
correlated\cite{LinAPL10}. The values of $n$ and $\mu$ that we
obtain here for the CR fall on the generic dependence found in
Ref. \onlinecite{LinAPL10}. CR was also observed in monolayer
epitaxial graphene on the silicon side of silicon
carbide\cite{WitowskiPRB10,CrasseeNatPhys11}, where similar
mobility of carriers was found\cite{CrasseeNatPhys11}.

\begin{table}
\begin{tabular}{lc}
  \hline\hline
  Transition & v$_{F}$ (m/s)\\ \hline
  LL$_{0\rightarrow1,a}$ & 1.02 $\times$ 10$^{6}$\\
  LL$_{0\rightarrow1,b}$ & 1.11 $\times$ 10$^{6}$\\
  LL$_{-1\rightarrow0}$ & 1.09 $\times$ 10$^{6}$\\
  LL$_{1\rightarrow2}$ & 1.01 $\times$ 10$^{6}$\\
  \hline\hline
\end{tabular}
\caption{Fermi velocities found for different LL transitions.}
\label{TabvF}
\end{table}

Next we discuss the field dependence of the LL transitions.
From $\sigma_{+}(\omega)$ and $\sigma_{-}(\omega)$ it is
particularly evident that the two electron like and one hole
like resonances described above have a similar magnetic field
dependence. The energies of all low-index LL transitions,
namely LL$_{1\rightarrow2}$ (Fig. \ref{Figfitresults2nd}a),
LL$_{0\rightarrow1,a}$, LL$_{0\rightarrow1,b}$ and
LL$_{-1\rightarrow0}$ (Fig. \ref{Figfitresults2nd}b), clearly
follow the square-root dependence on magnetic field typical of
massless Dirac fermions (Eq. (\ref{EqEn})), in accordance with
previous observations in similar epitaxial graphene
samples\cite{SadowskiPRL06} and exfoliated monolayer
flakes\cite{JiangPRL07,DeaconPRB07}. Fitting the field
dependence of the LL transition energies using Eq. (\ref{EqEn})
provides the Fermi velocities listed in Table \ref{TabvF}; they
show a spread of about 10\%. The difference between the Fermi
velocities found from the electron like LL$_{0\rightarrow1,b}$
and hole like LL$_{-1\rightarrow0}$ transitions is only about
2\%. However, the presence of three distinct inflection points
in $\sigma_{xy}(\omega)$ corresponding to the energies of the
three LL transitions, is a clear sign that the transition
energies and therefore the Fermi velocities are different.

The spectral weights of the LL$_{1\rightarrow2}$,
LL$_{0\rightarrow1,a}$, LL$_{0\rightarrow1,b}$ and
LL$_{-1\rightarrow0}$ transitions increase with magnetic field
(Fig. \ref{Figfitresults2nd}d and \ref{Figfitresults2nd}e).
This is in a qualitative agreement with an increase of the
number of states within each LL with magnetic
field\cite{SadowskiPRL06,GusyninPRL07}. However, a closer look
at the absolute values of the spectral weights reveals a strong
reduction with respect to the theoretical expectation for ideal
graphene monolayers, as discussed in Section \ref{Discussion}.

The scattering rate of the LL transitions is about 3 to 4 meV
and magnetic field independent (Fig. \ref{Figfitresults2nd}f
and \ref{Figfitresults2nd}h). Although in the case of the
cyclotron peak we extracted the mobility using the scattering
rate and the cyclotron frequency, in the quantum limit where
the transition energy is proportional to $\sqrt{B}$, the
quasi-classical Eq. (\ref{Eqmu}) will formally result in
diverging mobility values at low fields.

\subsection{Temperature dependence}\label{temperature dependence}

\begin{figure}
\includegraphics[width=8.5cm]{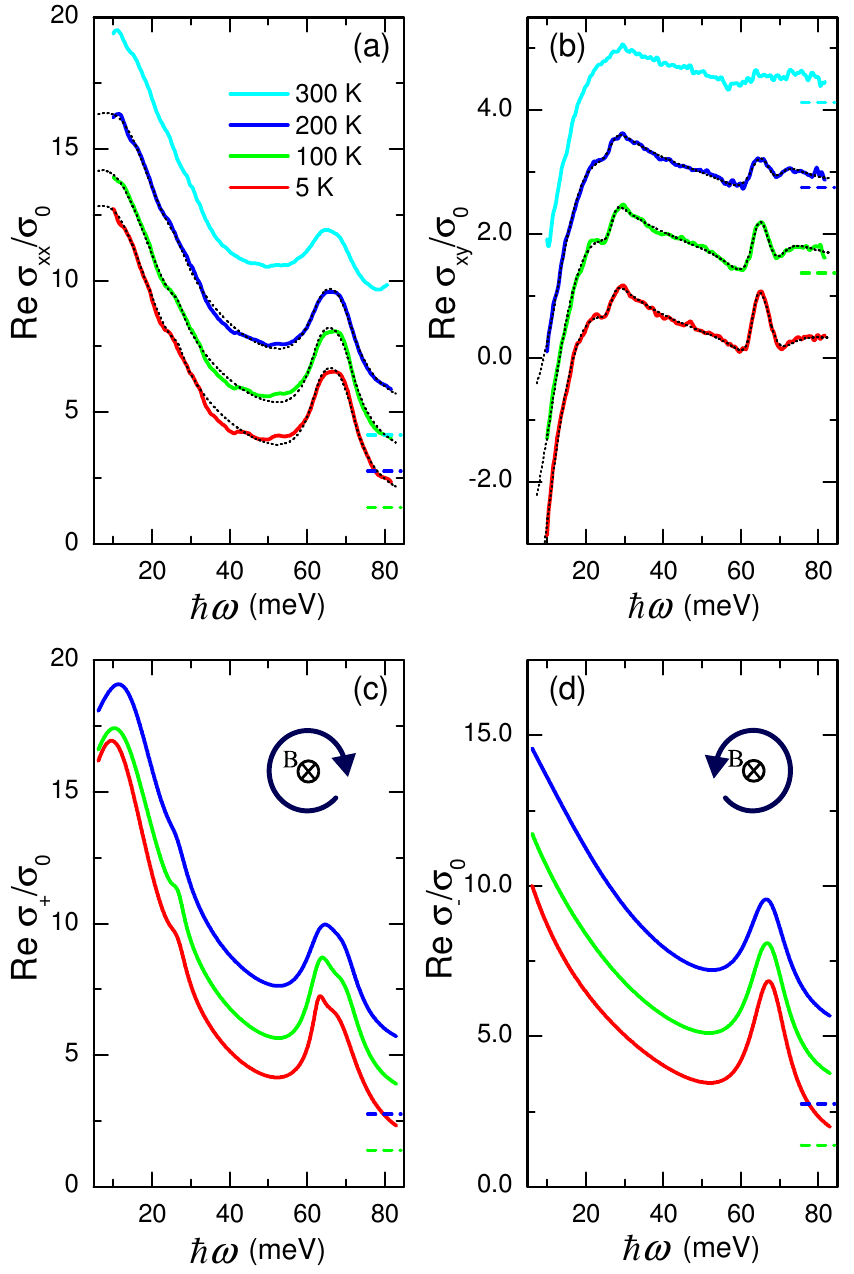}\\
\caption{
Magneto-optical conductivity of multilayer graphene at
3 T, normalized to the universal conductivity $\sigma_0$, for
several temperatures between 5 K and 300 K. The curves in all panels are
offset as indicated by the dashed lines. Larger offsets
correspond to higher temperatures. Panels (a) and (b) show
the measured spectra of $\sigma_{xx}(\omega)$ and
$\sigma_{xy}(\omega)$ (solid lines) and multi-component fits
(black dotted lines, only at 5, 100 and 200 K). In panels (c) and (d) the model-derived
$\sigma_{+}(\omega)$ and $\sigma_{-}(\omega)$ are shown.
} \label{Figtempdepend}
\end{figure}

The temperature dependence of $\sigma_{xx}(\omega)$ and
$\sigma_{xy}(\omega)$ at a fixed field of 3 T is shown in Fig.
\ref{Figtempdepend}a and \ref{Figtempdepend}b, respectively. In
both sets of curves, the CR structure changes weakly with
temperature. The situation is different for the LL transitions.
In $\sigma_{xx}(\omega)$ the LLs are preserved at all
temperatures, although at elevated temperatures the peaks are
about 3 meV broader than at low temperatures as obtained from
the fits shown in the same graphs. Their spectral weight is
almost temperature independent. However, in
$\sigma_{xy}(\omega)$, the spectral features corresponding to
the LL transitions are strongly diminished at high temperatures
and almost disappear at room temperature.

The extinction of the spectral structures corresponding to the
LL transitions in $\sigma_{xy}(\omega)$ with warming up is due
to the simultaneous presence of electron and hole like LL
transitions as seen in the graphs of $\sigma_{+}(\omega)$ and
$\sigma_{-}(\omega)$, plotted in Fig. \ref{Figtempdepend}c and
\ref{Figtempdepend}d. At high temperatures, the electron- and
hole- like components that have slightly different Fermi
velocities overlap more, due to the increased broadening. In
$\sigma_{xy}(\omega)$ the contributions are subtractive and
more overlap results in weaker spectral structures. On the
other hand, in $\sigma_{xx}(\omega)$, which is unsensitive to
the sign of the charge carriers, the contributions are additive
and the transition peak appears less affected.

\subsection{Effects of environmental doping}\label{environmental evolution}

\begin{figure}
\includegraphics[width=8.5cm]{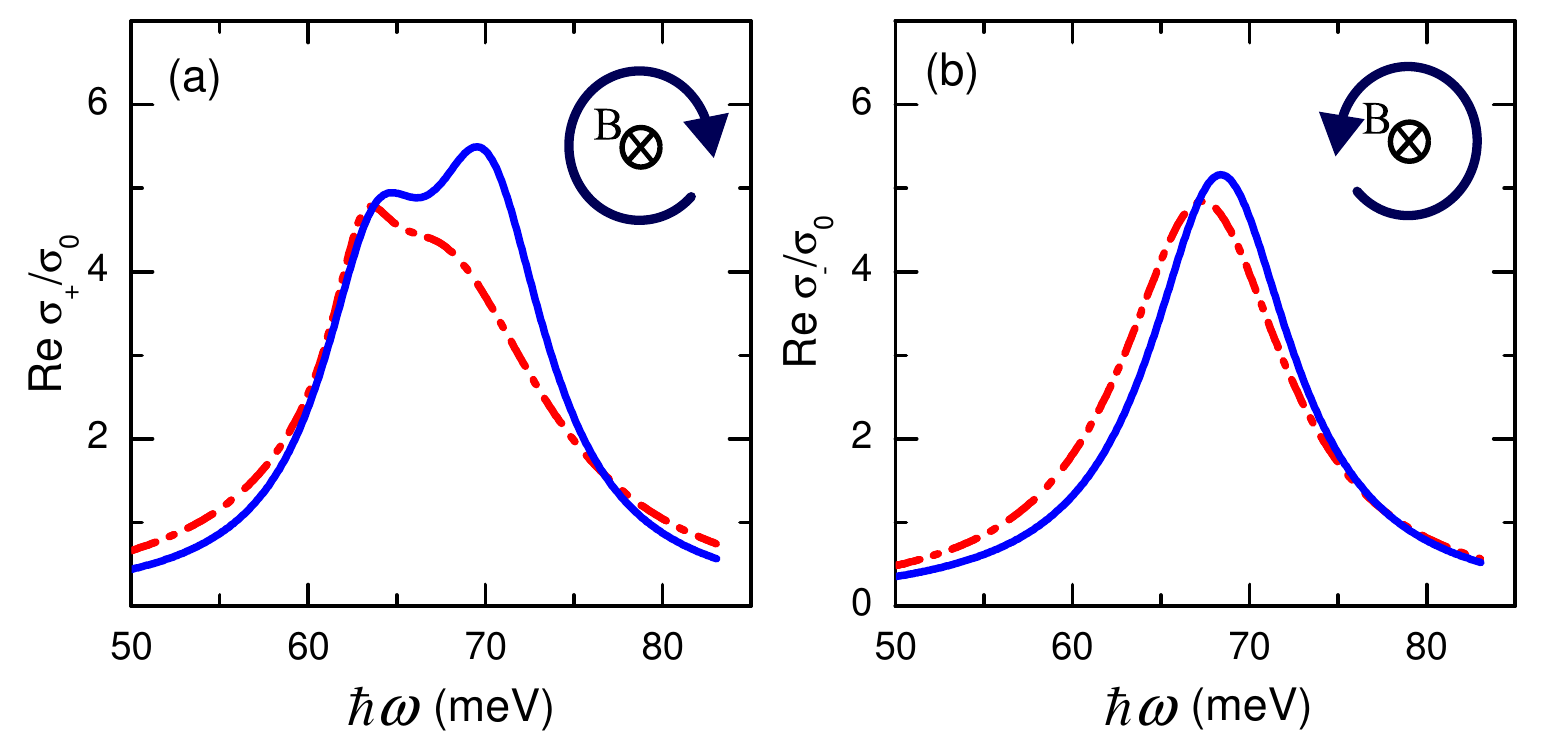}\\
\caption{The effect of environmental doping on the LL transitions at B = 3 T and T = 5 K.
Panel (a):  the contributions of the LL$_{0\rightarrow1,a}$
and LL$_{0\rightarrow1,b}$ transitions to $\sigma_{+}(\omega)$. Panel (b): the contribution of the LL$_{-1\rightarrow0}$ transition
to $\sigma_{-}(\omega)$. In both panels, the red dash-dotted lines
correspond to the measurement series as a function of
temperature at constant magnetic field (Fig.
\ref{Figtempdepend}), while the blue solid lines correspond to the
measurement series as a function of magnetic field at constant
temperature (Fig. \ref{Figmagneticfielddependence}). The curves are derived from the multi-component fits of experimental data.}
\label{FigB3Tcomparison}
\end{figure}

A careful inspection of the two series of measurements (Figs.
\ref{Figmagneticfielddependence} and \ref{Figtempdepend}) shows
that the spectra from the different series taken at the same
experimental conditions (5 K and 3 T) are not precisely the
same, especially close to the LL$_{0\rightarrow1,a}$,
LL$_{0\rightarrow1,b}$ and LL$_{-1\rightarrow0}$ transition
energies. As was mentioned above, between these measurements
the sample was held in dry air, therefore it is reasonable to
assume that the difference is due to the effect of the
environmental molecular contamination\cite{SchedinNatureMat07}.
In the case of multilayer graphene, the electronic properties
of the outmost layer are most strongly modified, although one
cannot exclude a certain effect on inner layers as well.

Fig. \ref{FigB3Tcomparison}a shows the contributions of the
LL$_{0\rightarrow1,a}$ and LL$_{0\rightarrow1,b}$ transitions
to $\sigma_{+}(\omega)$, for both measurements at 5 K and 3 T.
The low energy LL$_{0\rightarrow1,a}$ transition with $v_F$ =
1.02 $\times$ 10$^{6}$ m/s is unaltered. The high energy
LL$_{0\rightarrow1,b}$ transition with $v_F$ = 1.11 $\times$
10$^{6}$ m/s however shows a clear change: the structure is
much sharper in the second measurement. In Fig.
\ref{FigB3Tcomparison}b the contribution of the
LL$_{-1\rightarrow0}$ transition to $\sigma_{-}(\omega)$ is
plotted for both measurements. Similar to
LL$_{0\rightarrow1,b}$, it shows a sharpening in the second
measurement. The fact that both LL$_{0\rightarrow1,b}$ and
LL$_{-1\rightarrow0}$ transition peaks are changed with
environmental doping suggests that these transitions originate
from the bands in the outmost graphene layer, which is most
affected by surface contamination. The LL$_{0\rightarrow1,a}$
transition with the lowest $v_F$ is unchanged by the
environmental doping, therefore it probably comes from deeper
graphene layers.

Fig. \ref{Figfitresults2nd}e shows that the spectral weights of
the LL$_{0\rightarrow1,b}$ and LL$_{-1\rightarrow0}$
transitions are approximately equal, which indicates a balance
between electrons and holes in the top layer. The Fermi
velocity of the electrons (1.11 $\times$ 10$^{6}$ m/s) in that
layer is slightly larger than the one of the holes (1.09
$\times$ 10$^{6}$ m/s). Similar results were found for
monolayer exfoliated\cite{DeaconPRB07} and few layer
CVD\cite{LuicanPRL11} graphene. The electron-hole asymmetry in
our work (about 2\%) is much smaller than the one found in Ref.
\onlinecite{LuicanPRL11}, which might be due to different
rotation angles between the layers in the used samples. Indeed,
the asymmetry between electrons and holes was experimentally
shown to depend on the relative rotation between subsequent
graphene layers\cite{LuicanPRL11}, where a large (small)
rotation between the layers gives a small (large) asymmetry.

\section{Discussion}\label{Discussion}

\begin{figure}
\includegraphics[width=4cm]{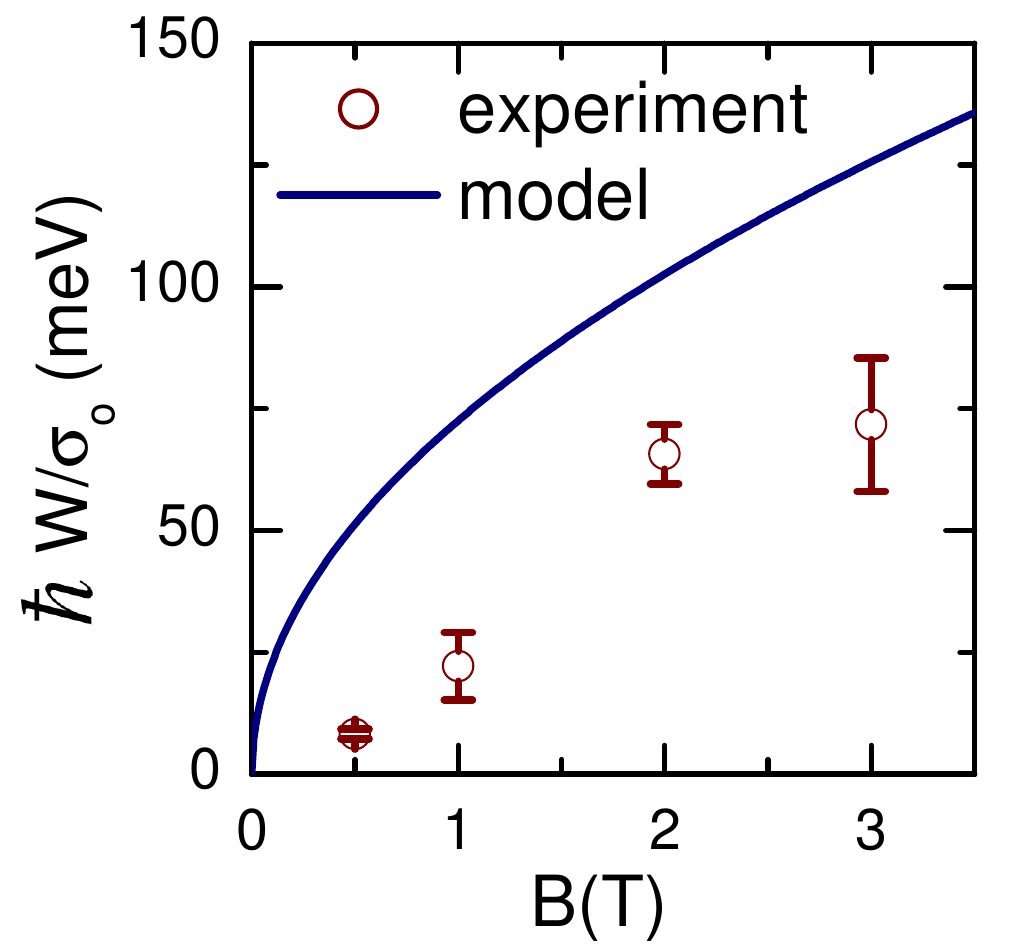}\\
\caption{Total spectral weight of the LL$_{0\rightarrow1,a}$,
LL$_{0\rightarrow1,b}$ and LL$_{-1\rightarrow0}$ transitions
(circles) as compared to the theoretical prediction for the
spectral weight of the sum of the LL$_{0\rightarrow1}$ and
LL$_{-1\rightarrow0}$ transitions in ideal monolayer
graphene (solid line) with the chemical potential between E$_{-1}$ and E$_{1}$.}
\label{FigSpectralWeight}
\end{figure}

Multilayer graphene on the C-side of SiC is often regarded as a
stack of twisted monolayers, electronically decoupled from each
other due to random rotational stacking, where the doping level
varies across the layers due to the effects of the substrate
and the surface contamination. Our observation of the CR and
the LL transitions (LL$_{1\rightarrow2}$, LL$_{0\rightarrow1}$,
LL$_{-1\rightarrow0}$) in the same spectra is a clear
indication of this doping variation. Indeed, since only
transitions between occupied and empty states can occur, their
simultaneous activation implies different positions of the
chemical potential with respect to the Dirac point energy.
However, the activation of different LL transitions may also be
caused by a spatial doping inhomogeneity.

The nearly perfect square-root dependence of the LL transition
energies on magnetic field found in previous
work\cite{SadowskiPRL06,OrlitaSemiSciTech10} and confirmed by
the present measurements (Fig.
\ref{Figmagneticfielddependence}a,b) is a signature of massless
Dirac dispersion inherent to monolayer graphene. The same field
dependence of LLs is obtained by STM
\cite{MillerScience09,SongNat10}. Accordingly, ARPES
measurements \cite{SprinklePRL09} show multiple Dirac cones
from individual layers. However, a number of our observations
is difficult to fit into a simple picture of completely
isolated monolayers. One of them is a significant (about 10\%)
spread of the Fermi velocity in the same sample. The largest
values $v_F$ = 1.09 - 1.11 $\times$ 10$^{6}$ m/s we attribute
to the outmost graphene layer, based on the effect of the
surface contamination, while the smallest values, $v_F$ = 1.01
- 1.02 $\times$ 10$^{6}$ m/s, correspond to the inner layers,
where the transitions LL$_{0\rightarrow1,a}$ and
LL$_{1\rightarrow2}$ are active due to a weak electron doping.
It was theoretically predicted\cite{LopesDosSantosPRL07,
LaissardiereNanoLett10, ShallcrossPRB10} that in rotationally
stacked graphene layers the effect of the interlayer
interaction is to reduce the Fermi velocity with respect to its
`bare' value in monolayer graphene. The variation of $v_F$ that
we find can thus be attributed to the effect of layer twisting
and a randomness of the rotation angles between various layers.

It is important to compare not only the energies but also the
optical spectral weights $W$ of the LL transitions with
theoretical expectations\cite{AbergelPRB07,GusyninPRL07}.
Experimentally, we obtain the weights from the multi-component
spectra fitting, described in Section \ref{multi component
fitting}. Let us consider the spectral weights of the
LL$_{0\rightarrow1}$ and LL$_{-1\rightarrow0}$ transitions.
Assuming that the chemical potential is between the first
electron and hole LLs ($E_{-1}$ and $E_{1}$), the total weight
of these transitions in monolayer graphene is given, according
to the Kubo formalism for non-interacting Dirac fermions, by
the transition energy itself \cite{GusyninPRL07}:

\begin{eqnarray}
\hbar W/\sigma_0 = 2(E_1-E_0) = 2\sqrt{2 e \hbar v_F^2 |B|}\label{EqW}
\end{eqnarray}

\noindent This dependence on magnetic field for $v_{F}$ = 1.0
$\times$ 10$^{6}$ m/s is plotted in Fig.
\ref{FigSpectralWeight} as a solid line. The symbols show the
sum of the experimental spectral weights of the
LL$_{0\rightarrow1,a}$ , LL$_{0\rightarrow1,b}$ and
LL$_{-1\rightarrow0}$ transitions. The experimental points are
clearly below the theoretical curve. From here it follows that
even if only in one graphene layer the chemical potential is
between $E_{-1}$ and $E_{1}$ then the total spectral weight
observed is about 2 times smaller than the theoretical
expectation. In reality it is likely that the number of layers
in the present sample that satisfy the condition for the
chemical potential is larger, which would make the deviation
even stronger.

A possibly related experimental observation is the presence of
the optical absorption background found from the fitting
results of the phenomenological cyclotron multi-component
model. This component, with vanishing $\omega_c$, has a
substantial spectral weight spread over a broad frequency
range. The existence of the background shows that a significant
amount of the charge carriers in the graphene layers neither
fall into well defined CR nor LL transitions and signals a
departure from the isolated monolayer description. In terms of
the optical sum rule, the missing spectral weight of the LL
transitions is transferred to the background. An intriguing
question is whether this transfer is caused by interlayer
coupling or by many-body effects within individual layers, such
as electron-electron and electron-phonon interactions.

One should notice that the mobility and density of carriers in
epitaxial graphene show a large variation from sample to
sample, even when they are prepared under similar conditions
\cite{LinAPL10}. This shows that the interlayer twist angle,
which is at the moment difficult to control experimentally, is
a crucial parameter affecting the electronic and therefore
optical properties of epitaxial graphene.

\section{Conclusions}\label{Conclusion}

Using magneto-optical infrared Hall spectroscopy, the charge
dynamics in multilayer epitaxial graphene grown on the C-side
of SiC was studied in magnetic fields up to 7 T. The diagonal
and the Hall conductivities were extracted from the absorption
and the Faraday rotation spectra, respectively. The latter is
sensitive to the sign of charge carriers that allowed us to
distinguish electrons and hole like transitions. The mobility
and charge density of electrons were found, which makes this
technique a useful contactless characterization tool.

A general multi-component model, Eq. (\ref{EqSpm2}), provided
excellent fits at each field to $\sigma_{xx}(\omega)$ and
$\sigma_{xy}(\omega)$ simultaneously. This analysis revealed
the coexistence of optical transitions between individual LLs
with a square-root dependence of the transition energies on
magnetic field as expected for isolated monolayer graphene and
a quasi-classical CR showing a linear magnetic field
dependence. This is a clear indication of the doping variation
across the layers.

We find the simultaneous presence of at least two distinct
peaks due to transitions between Landau levels 0 and 1
(electrons). The separation between these peaks corresponds to
a difference between the Fermi velocities of about 10\%. One
more peak due to a transition between LLs -1 (holes) and 0 is
observed. The effect of the surface contamination on the
spectra tells that both electrons and holes are present in the
top layer and that electrons have slightly higher Fermi
velocity than the holes (by 2\%). The variation of the Fermi
velocity is probably related to random twisting angles between
graphene layers.

The spectral weight of the LL transitions is shown to be
significantly reduced with respect to the theoretical
expectation for a stack of fully decoupled graphene monolayers,
assuming the picture of non-interacting electrons within each
layer. This missing spectral weight correlates with the
presence of an unexpected broadband optical absorption, which
is also inconsistent with this simplified theoretical model.

Although the transition energies of the LL transitions clearly
follow the magnetic field dependence expected for isolated
graphene, in order to come to a complete picture of the complex
electronic structure of multilayer graphene we need to
understand the variation of the Fermi velocity, the small
optical spectral weight of the LL transition and the broad
absorption background in relation to the twist angle of the
layers. Therefore, a systematic study of samples where this
angle is experimentally controlled, is required.

This work was supported by the Swiss National Science
Foundation (SNSF) by the grants 200021-120347 and IZ73Z0-128026
(SCOPES program), through the National Centre of Competence in
Research 'Materials with Novel Electronic Properties-MaNEP'. We
thank S.G. Sharapov for useful discussions.


\begin{thebibliography}{99}

\bibitem{BergerJPCB04} C. Berger, Z. Song, T. Li, X. Li, A. Y.
    Ogbazghi, R. Feng, Z. Dai, A. N. Marchenkov, E. H. Conrad,
    P. N. First, and W. A. de Heer, J. Phys.
    Chem. B, \textbf{108}, 19912 (2004).

\bibitem{BergerScience06} C. Berger, Z. Song, X. Li, X. Wu, N.
    Brown, c. Naud, D. Mayou, T. Li, J. Hass, A. N. Marchenkov,
    E. H. Conrad, P. N. First, and W. A. de Heer,
    Science, \textbf{312}, 1191 (2006).

\bibitem{EmtsevNatureMat09} K.V. Emtsev, A. Bostwick, K. Horn,
    J. Jobst, G. L. Kellogg, L. Ley, J. L. McChesney, T. Ohta,
    S. A. Reshanov, J. R\"{o}hrl, E. Rotenberg, A. K. Schmid,
    D. Waldmann, H. B. Weber, and Th. Seyller, Nature Mater. \textbf{8}, 203 (2009).

\bibitem{TzalenchukNatureNano10} A. Tzalenchuk, S. Lara-Avila,
    A. Kalaboukhov, S. Paolillo, M. Syv\"{a}j\"{a}rvi, R.
    Yakimova, O. Kazakova, T. J. B. M. Janssen, V. Fal'lko, and S. Kubatkin, Nature Nano. \textbf{5}, 186 (2010).

\bibitem{RiedlPRL09} C. Riedl, C. Coletti, T. Iwasaki, A.A.
    Zakharov, and U. Starke, Phys. Rev. Lett.
    \textbf{103}, 246804 (2009).

\bibitem{SpeckMatSciForum10} F. Speck, M. Ostler, J. R\"{o}hrl,
    J. Jobst, D. Waldmann, M. Hundhausen, L. Ley, H. B. Weber,
    and Th. Seyller, Mat. Sci. Forum \textbf{645}, 629-632 (2010).

\bibitem{FirstMRSBull10} P. N. First, W. A. de Heer, Th.
    Seyller, C. Berger, J. A. Stroscio, and J.-S. Moon, MRS
    Bull., \textbf{35}, 296 (2010).

\bibitem{HassPRL08}J. Hass, F. Varchon, J. E. Mill\'{a}n-Otoya,
    M. Sprinkle, N. Sharma, W. A. de Heer, C. Berger, P. N.
    First, L. Magaud, and E. H. Conrad, Phys. Rev. Lett. \textbf{100}, 125504 (2008).

\bibitem{SadowskiPRL06} M.L. Sadowski, G. Martinez, M.
    Potemski, C. Berger, and W. A. de Heer, Phys. Rev. Lett.
    \textbf{97}, 266405 (2006).

\bibitem{MillerScience09} D. L. Miller, K. D. Kubista, G. M.
    Rutter, M. Ruan, W. A. de Heer, P. N. First, and J. A.
    Stroscio, Science \textbf{324}, 924 (2009).

\bibitem{SongNat10}Y. J. Song, A. F. Otte, Y. Kuk, Y. Hu, D. B.
    Torrance, P. N. First, W. A. de Heer, H. Min, S. Adam, M.
    D. Stiles, A. H. MacDonald, and J. A. Stroscio, Nature \textbf{467}, 185 (2010).

\bibitem{SprinklePRL09} M. Sprinkle, D. Siegel, Y. Hu, J.
    Hicks, A. Tejeda, A. Taleb-Ibrahimi, P. Le F\`{e}vre, F.
    Bertran, S. Vizzini, H. Enriquez, S. Chiang, P. Soukiassian,
    C. Berger, W. A. de Heer, A. Lanzara, and E. H. Conrad,
    Phys. Rev. Lett. \textbf{103}, 226803 (2009).

\bibitem{WuAPL09}X. Wu, Y. Hu, M. Ruan, N. K. Madiomanana, J.
    Hankinson, M. Sprinkle, C. Berger, and W. A de Heer, Appl.
    Phys. Lett. \textbf{95}, 223108 (2009).

\bibitem{GuineaPRB06}F. Guinea, A. H. Castro Neto, and N. M. R.
    Peres, Phys. Rev. B \textbf{73}, 245426 (2006).

\bibitem{mcCannPRL06}E. McCann and V. I. Fal'ko, Phys. Rev.
    Lett. \textbf{96}, 086805 (2006).

\bibitem{LopesDosSantosPRL07}J. M. B. Lopes dos Santos, N. M.
    R. Peres, and A. H. Castro Neto, Phys. Rev. Lett.
    \textbf{99}, 256802 (2007).

\bibitem{ShallcrossPRL08} S. Shallcross, S. Sharma, and O. A.
    Pankratov, Phys. Rev. Lett. \textbf{101}, 056803 (2008).

\bibitem{LaissardiereNanoLett10} G. T. de Laissardi\`{e}re,
    D. Mayou, and L. Magaud, Nano Lett. \textbf{10},
    804 (2010).

\bibitem{ShallcrossPRB10} S. Shallcross, S. Sharma, E.
    Kandelaki, and O. A. Pankratov, Phys. Rev. B \textbf{81},
    165105 (2010).

\bibitem{MelePRB01} E. J. Mele, Phys. Rev. B \textbf{81},
    161405 (2010).

\bibitem{BistritzerPRB10} R. Bistritzer and A. H. MacDonald,
   \emph{Phys. Rev. B} \textbf{81}, 245412 (2010).

\bibitem{BistritzerArxiv11} R. Bistritzer and A. H. MacDonald,
   ArXiv:1101.2606 (2011).

\bibitem{CrasseeNatPhys11} I. Crassee, J. Levallois, A. L.
    Walter, M. Ostler, A. Bostwick, E. Rotenberg, Th. Seyller,
    D. van der Marel, and A. B. Kuzmenko,
    Nature Physics \textbf{7}, 48-51 (2011).

\bibitem{AbergelPRB07} D. S. L. Abergel and V. I. Fal'ko,
    Phys. Rev. B \textbf{75}, 155430 (2007).

\bibitem{GusyninJPCM07} V. P. Gusynin, S. G. Sharapov, and J.
    P. Carbotte, \emph{J. Phys. Condens. Matter} \textbf{19},
    026222 (2007).

\bibitem{OrlitaSemiSciTech10} M. Orlita and M. Potemski,
    Semicond. Sci. Technol. \textbf{25}, 06001 (2010).

\bibitem{text}Note that also transitions between electron-
    and hole- like LLs are optically allowed. However, for the used
    magnetic fields, these transitions are beyond the
    experimental spectral range, except at 0.5 T, where rather
    weak contributions from LL$_{-1\rightarrow2}$ and
    LL$_{-2\rightarrow 1}$ could be discerned.

\bibitem{AndoJPSJ02} T. Ando, Y. Zheng, and H. Suzuura,
    J. Phys. Soc. Jpn. \textbf{71}, 1318–1324 (2002).

\bibitem{GusyninNJP09} V. P. Gusynin, S. G. Sharapov, and J. P.
    Carbotte, New J. Phys. \textbf{11}, 095013 (2009).

\bibitem{LinAPL10} M. Y. Lin, C. Dimitrakopoulos, D. B. Farmer,
    S.-J. Han, Y. Wu, W. Zhu, D. K. Gaskill, J. L. Tedesco, R.
    L. Myers-Ward, C. R. Eddy, A. Grill, and P. Avouris, Appl. Phys. Lett. \textbf{97}, 112107 (2010).

\bibitem{WitowskiPRB10} A. M. Witowski, M. Orlita, R.
    St\c{e}pniewski, A. Wysmo{\l}ek, J. M. Baranowski, W.
    Strupi\'{n}ski, C. Faugeras, G. Martinez, and M. Potemski,
    Phys. Rev. B \textbf{82}, 165305 (2010).

\bibitem{JiangPRL07} Z. Jiang, E. A. Henriksen, L. C. Tung,
    Y.-J. Wang, M. E. Schwartz, M. Y. Han, P. Kim, and H. L.
    Stormer, Phys. Rev. Lett. \textbf{98}, 197403 (2007).

\bibitem{DeaconPRB07} R. S. Deacon, K.-C. Chuang, R. J.
    Nicholas, K. S. Novoselov, and A. K. Geim, Phys. Rev. B
    \textbf{76}, 081406(R) (2007).

\bibitem{GusyninPRL07} V. P. Gusynin, S. G. Sharapov, and J. P.
    Carbotte, Phys. Rev. Lett. \textbf{98} 157402 (2007).

\bibitem{SchedinNatureMat07} F. Schedin, A. K. Geim, S. V.
    Morozov, E. W. Hill, P. Blake, M. I. Katsnelson, and K. S.
    Novoselov,
     Nature Mat. \textbf{6}, 652-655 (2007).

\bibitem{LuicanPRL11} A. Luican, G. Li, A. Reina, J. Kong, R.
    R. Nair, K. S. Novoselov, A. K. Geim, and E. Y. Andrei,
    Phys. Rev. Lett. \textbf{106}, 126802 (2011).

\end{thebibliography}
\end{document}